\documentstyle[12pt]{article}
\newcommand{\es}{\\[2mm]}

\newcommand{\journal}[4]{{\em #1~}#2\,(19#3)\,#4;}

\newcommand{\ijmp}{\journal {Int. J. Mod. Phys.}}

\newcommand{\pr}{\journal {Phys. Rev.}}

\newcommand{\jmp}{\journal {J. Math. Phys.}}

\newcommand{\cmp}{\journal {Comm. Math. Phys.}}
\newcommand{\cqg}{\journal {Class. Quantum Grav.}}

\newcommand{\np}{\journal {Nucl. Phys.}}
\newcommand{\pl}{\journal {Phys. Lett.}}

\newcommand{\nc}{\journal {Nuovo Cim.}}

\makeatletter
\@addtoreset{equation}{section}
\makeatother

\def\Lp{\displaystyle{\biggl(}}
\def\Rp{\displaystyle{\biggr)}}
\def\LP{\displaystyle{\Biggl(}}
 \def\wti{\widetilde}
\def\RP{\displaystyle{\Biggr)}}
\newcommand{\lp}{\left(}\newcommand{\rp}{\right)}
\newcommand{\lc}{\left[}\newcommand{\rc}{\right]}
\newcommand{\lac}{\left\{}\newcommand{\rac}{\right\}}
\newcommand{\G}{\Gamma}
\newcommand{\D}{\Delta}

\renewcommand{\b}{\beta}
\renewcommand{\d}{\delta}

\newcommand{\f}{\phi}
\newcommand{\g}{\gamma}

 \renewcommand{\L}{\Lambda}

\newcommand{\n}{\nu}

 \renewcommand{\O}{\Omega}

\newcommand{\r}{\rho}
 \renewcommand{\S}{\Sigma}

\renewcommand{\AA}{{\cal A}}
\newcommand{\BB}{{\cal B}}

\newcommand{\FF}{{\cal F}}

\newcommand{\HH}{{\cal H}}

\newcommand{\MM}{{\cal M}}
\newcommand{\NN}{{\cal N}}
\newcommand{\OO}{{\cal O}}

\newcommand{\QQ}{{\cal Q}}
\newcommand{\RR}{{\cal R}}
\newcommand{\SS}{{\cal S}}
\newcommand{\TT}{{\cal T}}
\newcommand{\UU}{{\cal U}}
\newcommand{\VV}{{\cal V}}
\newcommand{\WW}{{\cal W}}

\newcommand{\complex}{{\kern .1em {\raise .47ex
\hbox {$\scriptscriptstyle |$}}
    \kern -.4em {\rm C}}}
\newcommand{\real}{{{\rm I} \kern -.19em {\rm R}}}
\newcommand{\rational}{{\kern .1em {\raise .47ex
\hbox{$\scripscriptstyle |$}}
    \kern -.35em {\rm Q}}}
\renewcommand{\natural}{{\vrule height 1.6ex width
.05em depth 0ex \kern -.35em {\rm N}}}
\newcommand{\dint}{\displaystyle{\int}}
\newcommand{\xint}{\dint d^2 \! x \, }

\newcommand{\pa}{\partial}
\newcommand{\pad}[2]{{\frac{\partial #1}{\partial #2}}}
\newcommand{\fud}[2]  {{\displaystyle{\frac{\delta #1}{\delta #2}}}}
\newcommand{\dfrac}[2]{{\displaystyle{\frac{#1}{#2}}}}
\newcommand{\dsum}[2]{\displaystyle{\sum_{#1}^{#2}}}

\newcommand{\sla}{\raise.15ex\hbox{$/$}\kern -.57em}

\newcommand{\twiddle}{\lower.9ex\rlap{$\kern -.1em\scriptstyle\sim$}}

\renewcommand{\pad}[2]{{\displaystyle{\frac{\partial #1}{\partial #2}}}}
\newcommand{\equ}[1]{(\ref{#1})}

\newcommand{\eq}{\begin{equation}}
\newcommand{\eqn}[1]{\label{#1}\end{equation}}
\newcommand{\eea}{\end{eqnarray}}
\newcommand{\eqa}{\begin{eqnarray}}
\newcommand{\eqan}[1]{\label{#1}\end{eqnarray}}
\newcommand{\ba}{\begin{array}}
\newcommand{\ea}{\end{array}}
\newcommand{\eqac}{\begin{equation}\begin{array}{rcl}}
\newcommand{\eqacn}[1]{\end{array}\label{#1}\end{equation}}

\begin{document}


{\ }

\vspace{20mm}

\vspace{2cm}
\centerline{\LARGE Algebraic characterization of anomalies}\vspace{2mm}
\centerline{\LARGE in chiral $\WW_3-$gravity}

\vspace{9mm}

\centerline{ Marcelo Carvalho, Luiz Claudio Queiroz Vilar}
\centerline{ and }
\centerline{S.P. Sorella}
\vspace{4mm}
\centerline{\it C.B.P.F}
\centerline{\it Centro Brasileiro de Pesquisas Fisicas,}
\centerline{\it Rua Xavier Sigaud 150, 22290-180 Urca}
\centerline{\it Rio de Janeiro, Brazil}
\vspace{10mm}

\centerline{{\normalsize {\bf PACS: 11.15.Bt }} }
\vspace{4mm}

\centerline{{\normalsize {\bf REF. CBPF-NF-042/94}} }

\vspace{4mm}
\vspace{10mm}

\centerline{\Large{\bf Abstract}}\vspace{2mm}
\noindent
The anomalies which occur in chiral $\WW_{3}-$gravity are characterized
by solving the $BRS$ consistency condition.

\setcounter{page}{0}
\thispagestyle{empty}

\vfill
\pagebreak

%
\section{Introduction}

$\WW-$algebras~\cite{zam} are an extension of the Virasoro
algebra. They describe the commutation relations between the
components of the
stress-energy tensor $(T_{++}, T_{--})$ and currents
$(W_{++++\cdot \cdot \cdot},W_{----\cdot \cdot \cdot})$
of higher spin (see ref.~\cite{trst} for a general
introduction).

Among the various $\WW-$algebras considered in the recent
literature, the so called $\WW_3-$algebra
plays a rather special role, due to the fact that it has a simple
field theory realization. The corresponding field model, known as
$\WW_{3}-$gravity, yields a generalization of the usual bosonic
string action. It is available in a chiral~\cite{hl} as
well as in a nonchiral version~\cite{nonchiral}. The starting field
content of the model is given by a set of free scalar fields
$\f^{i} (i=1,...,D)$ which are used to study the current algebra of the
spin-two and spin-three operators  $T_{++}$ (resp. $T_{--}$)
and  $W_{+++}$  (resp. $W_{---}$)
\eq
    T_{++}=\frac{1}{2}\pa_{+}\f^{i}\pa_{+}\f^{i} \ , \qquad
W_{+++}=\frac{1}{3}d_{ijk}\pa_{+}\f^{i}\pa_{+}\f^{j}\pa_{+}\f^{k}  \ .
\eqn{t-w-operators}
The quantities $d_{ijk}$ are totally symmetric and chosen to satisfy
\eq
 d_{(ij}^{m} d_{k)lm}= \d _{(ij}\d_{k)l} \ .
\eqn{first-d-tensor}
The operators in \equ{t-w-operators} can be included in the initial
free scalar action by
coupling them to external fields  $h_{--},B_{---}$
(resp. $h_{++},B_{+++}$).
It is a remarkable fact then that the resulting action
exhibits a set of local invariances which can be gauge fixed using the
Batalin-Vilkovisky~\cite{bv} procedure.
The model is thus constrained by a classical
Slavnov-Taylor identity
which can be taken as the starting point for the analysis of
the quantum properties
of the $(T-\WW)-$current algebra. As it is well known, the existence of
anomalies in the quantum extension of the Slavnov-Taylor identity
turns out to be related to the presence of central charges in the
corresponding current algebra.

This is the case of the $\WW_{3}-$gravity, for
which several types of anomalies have been found~\cite{hl,nonchiral}.
Let us recall indeed that, besides the gravitational anomalies
which depend on the fields $h_{--},B_{---}$, the model
possesses a pure matter field anomaly whose origin lies in the
nonlinearity of the $\WW_{3}-$algebra.

Let us also remark that these anomalies have been essentially
detected by a direct computation of the relevant Feynman graphs
which contribute to the one and two loop effective action.
However, up to our knowledge, a purely algebraic characterization
based on the cohomological analysis of the $BRS$ consistency condition
is still lacking. The aim of the present work is to fill this gap.

In order to avoid too many technical details we shall limit here
to present the analysis of the chiral $\WW_{3}-$models
for which the $d_{ijk}$'s in
equations\equ{t-w-operators}, \equ{first-d-tensor}
are traceless~\footnote{Conditions \equ{first-d-tensor} and \equ{traceless}
define a nonempty set of algebras, known as the magic Jordan
algebras~\cite{magic}.}, i.e.
\eq
 d^{j}_{ij}=0 \ .
\eqn{traceless}
The generalization to the nontraceless case as well as to the nonchiral
models can be done in the same way and does not present any additional
complication.

The paper is organized as follows. In Sect. 2 the
quantization of chiral $\WW_{3}-$gravity is shortly recalled.
In Sect. 3 we discuss the
$BRS$ consistency condition for the anomalies and we derive the
corresponding descent equations. The final part of this section is
devoted to the introduction of an operator $\d$ which allows to decompose
the exterior space-time derivative as a $BRS$ commutator~\cite{silvio}.
This important feature will provide a powerful method for solving
the descent equations.

In Sect. 4 we present the computation
of the cohomology of the linearized Slavnov-Taylor operator.
In particular,
we will be able to establish an useful formula, called here the
{\it Russian-like}
formula, which will greatly simplify the full algebraic analysis.
This formula denotes a special combination of the ghost fields which allows
to compute in a very simple and elegant way all the relevant cohomology
classes. The {\it Russian-like} formula is one of the main results of the
present paper. It will reveal a simple and so far unnoticed
elementary structure of the $\WW_{3}-$gravity.

In Sect. 5 the anomalies which
affect the quantum extension of the Slavnov-Taylor identity are given.
Sect. 6 deals with some interesting property of the $\WW_{3}-$action.
We will see indeed that,
in complete analogy with the case of the bosonic string~\cite{wien},
the complete action of the $\WW_{3}-$gravity turns out to be cohomologically
trivial, suggesting then a topological interpretation of the model.

For the sake of clarity we have collected the most lengthy expressions
in two Appendices, respectively {\it App. A} and {\it App. B}.
Finally, in order to make contact with the two loop computation of
ref~\cite{hl}, the {\it App. C} contains a detailed discussion of the $BRS$
consistency condition at the order $\hbar^2$.

\section{The model and its quantization}

In this section we briefly review the quantization of chiral
$\WW_3-$gravity. Following~\cite{hl}  the gauge-fixed action we start with
is
\eq
S=S_{inv} +S_{gh} \ ,
\eqn{act}
where
\eq
S_{inv}= \xint \Lp \frac{1}{2}\pa_{+}\f^{i} \pa_{-}\f^{i} -
h_{--}T_{++} - B_{---}W_{+++} \Rp  \ ,
\eqn{so-action}
$T_{++}$ and $W_{+++}$ denoting respectively the operators
\eq
    T_{++}=\frac{1}{2}\pa_{+}\f^{i}\pa_{+}\f^{i} \ , \qquad
W_{+++}=\frac{1}{3}d_{ijk}\pa_{+}\f^{i}\pa_{+}\f^{j}\pa_{+}\f^{k}  \ .
\eqn{t-w}
They describe the coupling of a set of scalar fields $\f^{i} (i=1,...,D)$
to external gauge fields  $(h_{--}, B_{---})$~\cite{hl}. The quantities
$d_{ijk}$ in
\equ{t-w} are totally symmetric and satisfy the constraints
\eq
 d_{(ij}^{m} d_{k)lm}= \d _{(ij}\d_{k)l} \ , \qquad d^{j}_{ij}=0 \ .
\eqn{d-tensor}
The gauge-fixing term $S_{gh}$ of \equ{act}, using a Landau type
background gauge~\cite{beltrami}, reads
\eq\ba{cl}
 & S_{gh} = \xint s\,(b_{++}h_{--} + \beta_{+++} B_{---}) \\
    &{\ } =-\xint b_{++} \Lp \pa_{-} c_{-} + \pa_{+}h_{--} c_{-}
                           - h_{--} \pa_{+} c_{-}
   + 2 (\, \pa_{+}B_{---}\g_{--}-B_{---}\pa_{+}\g_{--})\,T_{++} \Rp \\
    & {\ }{\ }{\ }{\ }-\xint \beta_{+++} \Lp \pa_{-}\g_{--}
   + 2 \pa_{+}h_{--}\g_{--} - h_{--}\pa_{+}\g_{--}
     - 2B_{---}\pa_{+}c_{-} + \pa_{+}B_{---} c_{-}
    \Rp \ ,
\ea\eqn{sgf}
where $(c_{-},\g_{--})$ and $(b_{++},\beta_{+++})$ are respectively a
pair of ghost and antighost fields and $s$ denotes the $BRS$ operator
whose action on the fields is specified by
\eq\ba{l}
 s c_{-}   =  c_{-}\pa_{+}c_{-} +2 \g_{--}\pa_{+}\g_{--}T_{++} \ , \es
 s \g_{--} =  c_{-}\pa_{+}\g_{--} - 2\pa_{+} c_{-} \g_{--} \ ,     \es
 s \f^{i}  =  c_{-} \pa_{+}\f^{i} + \g_{--}d^{i}_{jk}
             \pa_{+}\f^{j}\pa_{+} \f^{k}  + 2 b_{++}\g_{--}
             \pa_{+}\g_{--} \pa_{+}\f^{i} \ ,                      \es
 s h_{--} =  \pa_{-} c_{-} + \pa_{+}h_{--} c_{-} - h_{--}\pa_{+} c_{-}
            + 2 (\pa_{+}B_{---} \g_{--}-
                  B_{---}\pa_{+}\g_{--}) T_{++} \ ,                \es
 s B_{---}= \pa_{-}\g_{--} + 2 \pa_{+}h_{--} \g_{--}
            - h_{--}\pa_{+}\g_{--}
            - 2B_{---}\pa_{+}c_{-}  + \pa_{+}B_{---} c_{-} \ ,     \es
 s b_{++} =  s \beta_{+++}=0  \ ,
\ea\eqn{brs}
and
\eq
      s S = 0 \ .
\eqn{brsinv}
Let us remark that the above transformations are nilpotent only
{\it on-shell}, i.e.
\eq\ba{l}
s^{2} h_{--}=-2 \g_{--} \pa_{+}\g_{--}\pa_{+}\f^{i}\fud{S}{\f^i} \ ,  \es
s^{2} \f^{i}= 2 \g_{--}\pa_{+}\g_{--}\pa_{+}\f^{i} \fud{S}{h_{--}} \ ,
\ea\eqn{on-shell}
and
\eq
s^{2} \lp  c_{-},{\ }\g_{--},{\ }B_{---},{\ }b_{++} \rp  = 0 \ .
\eqn{off-shell}
However, coupling the nonlinear $BRS$ transformations of
$(\f^{i},c_{-},\g_{--})$ in \equ{brs} to external
fields $(Y^{i}, \tau_{+}, \rho_{++})$
\eq
S_{ext}=\xint \Lp Y^{i}s \f^{i} + \tau_{+} s c_{-}
        + \rho_{++} s \g_{--} \Rp  \ ,
\eqn{extaction}
and making use of the indentities
\eq
  s h_{--} = - \fud{S}{b_{++}} \ , \qquad
  s B_{---} = -\fud{S}{\beta_{+++}} \ ,
\eqn{alg-eq}
one easily verifies that the total action $\S$
\eq
     \S = S_{inv} + S_{gh} + S_{ext} \ ,
\eqn{tot-act}
obeys the classical Slavnov-Taylor identity~\cite{hl}
\eq
     \SS(\S) = 0 \ ,
\eqn{slavnov}
with
\eq
  \SS(\S) = \xint \Lp \fud{\S}{Y^{i}} \fud{\S}{\f^{i}}
                   +  \fud{\S}{\tau_{+}} \fud{\S}{c_{-}}
                   +  \fud{\S}{\rho_{++}} \fud{\S}{\g_{--}}
                   -  \fud{\S}{b_{++}} \fud{\S}{h_{--}}
                   -  \fud{\S}{\beta_{+++}} \fud{\S}{B_{---}} \Rp  \ .
\eqn{slav-op}
As in the case of the bosonic string~\cite{beltrami}, identity \equ{slavnov}
is the starting point for studying the properties of the Green's functions
of the model with insertion of the composite operators
$(T_{++},W_{+++})$, i.e. \equ{slavnov} yields an algebraic
characterization of the $(T-W)$-current algebra.

Introducing now the linearized Slavnov-Taylor operator $\BB_{\S}$
\eq\ba{cl}
 \BB_{\S}  = \xint \Lp & \fud{\S}{Y^{i}} \fud{{\ }}{\f^{i}}
              + \fud{\S}{\f^{i}} \fud{{\ }}{Y^{i}}
              + \fud{\S}{\tau_{+}} \fud{{\ }}{c_{-}}
              + \fud{\S}{c_{-}} \fud{{\ }}{\tau_{+}}
              + \fud{\S}{\rho_{++}} \fud{{\ }}{\g_{--}}
              + \fud{\S}{\g_{--}} \fud{{\ }}{\rho_{++}} \\[4mm]
  &           - \fud{\S}{b_{++}} \fud{{\ }}{h_{--}}
              - \fud{\S}{h_{--}} \fud{{\ }}{b_{++}}
              - \fud{\S}{B_{---}} \fud{{\ }}{\beta_{+++}}
              - \fud{\S}{\beta_{+++}} \fud{{\ }}{B_{---}} \Rp \ ,
\ea\eqn{lin-slav}
one gets the nonlinear algebraic relation
\eq
        \BB_{\FF} \SS(\FF) = 0 \ ,
\eqn{alg-rel}
valid for an arbitrary functional $\FF$ with even ghost number. In particular
if $\FF$ is a solution of \equ{slavnov} then the corresponding linearized
operator $\BB_{\FF}$ is nilpotent, i.e. $\BB_{\FF} \BB_{\FF}=0$. Therefore
\eq
    \BB_{\S} \BB_{\S}=0 \ .
\eqn{nilp}

Let us conclude this section by displaying the ghost number $N_g$ and the
dimension of all the fields and sources
\begin{table}[hbt]
\centering
\begin{tabular}{|c|c|c|c|c|c|c|c|c|c|c|c|} \hline
 & $\f^{i}$ & $h_{--}$ & $B_{---}$ & $c_{-}$ & $\g_{--}$ & $b_{++}$
 & $\b_{+++}$
 & $Y^{i}$ & $\tau_{+}$ & $\r_{++}$  & $\BB_{\S}$ \\ \hline
$N_{g}$ & 0 & 0 & 0 & 1 & 1 & -1 & -1  & -1 & -2 & -2  & 1\\ \hline
$dim$ & 0 & 0 & -1 & 0 & -1 & 1 & 2  & 1 & 1 & 2  & 1\\ \hline
\end{tabular}
\caption[t1]{Ghost numbers and dimensions.}
\label{gh1-number-dim}
\end{table}


\section{The consistency condition and the descent equations}

At the quantum level the classical action \equ{tot-act} gives rise to
an effective action
\eq
    \G = \S + O{(\hbar)} \ ,
\eqn{eff-act}
which obeys the broken Slavnov-Taylor identity
\eq
\SS (\G)=  \hbar \AA \cdot\G  \ ,
\eqn{qslavnov}
where the insertion $\AA\cdot\G$ represents the possible breaking induced
by the radiative corrections. According to the
Quantum Action Principle~\cite{qap}
the lowest order nonvanishing contribution to the breaking
\eq
  \AA\cdot\G = \AA+ O(\hbar\AA) \ ,
\eqn{first-order}
is an integrated local functional of the fields and their derivatives with
dimension three and ghost number one which obeys the consistency condition
\eq
 \BB_{\S}\AA =0  \ .
\eqn{wz}
The latter is the first-order expansion of the exact relation
\eq
 \BB_{\G} \AA \cdot \G = 0 \ ,
\eqn{ex-rel}
which easily follows from equation \equ{alg-rel}, i.e. from
$\BB_{\G} \SS(\G) = 0$.

Setting $\AA = \int \AA_{2}^{1}$, condition \equ{wz} yields the
nonintegrated equation~\footnote{We adopt here the usual convention
of denoting with $\AA^{p}_{q}$ a q-form with ghost number p.},
\eq
   \BB_{\S} \AA_{2}^{1} + d \AA_{1}^{2} = 0 \ ,
\eqn{loc-wz}
where $\AA_{1}^{2}$ is a local functional with ghost number two and form
degree one and $d = dx^{+}\pa_{+} + dx^{-}\pa_{-}$ denotes the exterior
space-time derivative which, together with the linearized
operator $\BB_{\S}$,
obeys~\footnote{The operators $\BB_{\S}$ and $d$ increase respectively
the ghost number and the form degree by one unit.}
\eq
    \BB_{\S} d + d \BB_{\S} = d^2 =  \BB_{\S} \BB_{\S} = 0 \ .
\eqn{nilp-d}
A solution $\AA_{2}^{1}$ of eq. \equ{loc-wz} is said nontrivial if
\eq
   \AA_{2}^{1} \ne \BB_{\S} {\hat \AA_{2}^{0}} + d {\hat \AA_{1}^{1}} \ ,
\eqn{non-triv-wz}
with ${\hat \AA_{2}^{0}}$ and ${\hat \AA_{1}^{1}}$ local functionals of the
fields. In this case the integral of $\AA_{2}^{1}$  on space-time,
$\int \AA_{2}^{1}$, identifies a cohomology class of the operator
$\BB_{\S}$ in the sector of the integrated local functionals with
ghost number one. As it is well known this
corresponds to the appearence of an anomaly in the quantum
extension of the Slavnov-Taylor identity \equ{slavnov}.

Equation \equ{loc-wz}, due to the relations \equ{nilp-d} and
to the algebraic Poincar{\'e} Lemma~\cite{poincare}, generates a tower
of descent equations
\eq\ba{l}
    \BB_{\S} \AA_{2}^{1} + d \AA_{1}^{2} = 0 \ , \es
    \BB_{\S} \AA_{1}^{2} + d \AA_{0}^{3} = 0 \ , \es
    \BB_{\S} \AA_{0}^{3} = 0 \ ,
\ea\eqn{desc-eq}
with $\AA_{0}^{3}$ a local functional with ghost number three and form degree
zero. Let us also remark that, due to
Table~\ref{gh1-number-dim} and to the fact that the space-time
derivatives $(\pa_{-},\pa_{+})$ are of dimension one, all the cocycles
$\AA_{2}^{1}, \AA_{1}^{2}, \AA_{0}^{3}$ in \equ{desc-eq}
have dimension three. The task of the next sections will be that of finding
the most general local nontrivial solution of the descent equations
\equ{desc-eq}, giving thus an algebraic characterization of the anomalies
which can arise at the quantum level.

\subsection{Decomposition of the exterior derivative}

In order to solve the ladder \equ{desc-eq} we follow the algebraic setup
proposed by one of the authors~\cite{silvio} and successfully applied
to the study of the Yang-Mills~\cite{tatar} and of the
gravitational anomalies~\cite{grav}. The method is based on the introduction
of an operator $\d$ which decomposes the exterior derivative $d$ as
a $\BB_{\S}$ commutator,
i.e.
\eq
     d = - \lc \BB_{\S}, \d \rc \ .
\eqn{dec}
One easily verifies that, once the decomposition \equ{dec} has
been found, successive applications of the operator $\d$ on the
cocycle $\AA_{0}^{3}$ which solves the last equation of
the tower \equ{desc-eq} yield an explicit nontrivial solution for the higher
cocycles $\AA_{2}^{1}, \AA_{1}^{2}$. It is interesting to observe
that, actually, the decomposition \equ{dec} also occurs in the
topological field
theories~\cite{topol-dec} and in the string theory~\cite{wien}.

Let us emphasize that solving
the last equation of \equ{desc-eq} is a problem of local cohomology
instead of a modulo $d$ one. One sees thus that, due to the operator
$\d$, the characterization of the cohomology of $\BB_{\S}$ modulo $d$ is
essentially reduced to a problem of local cohomology. It is also worth to
recall that the latter can be systematically studied by using several
methods as, for instance, the spectral sequences
technique~\cite{dixon,b3cd}.

To find the decomposition \equ{dec} let us begin by specifying
the local functional space $\VV$ the operators $\BB_{\S}$ and $d$ act upon.
Taking into account that the classical action \equ{tot-act} is
invariant under
the constant shift symmetry $(\f^{i} \rightarrow \f^{i} + const)$ and
that, due to the couplings with the fields $(h_{--},\,B_{---})$
(see eq.\equ{so-action}), the matter fields $\f^{i}$ enter into the effective
action $\G$ with at least one derivative  $\pa_{+}$, it follows that
the space $\VV$ can be identified as the space of the local formal
power series in the variables ($h_{--}$, $B_{---}$, $c_{-}$, $\g_{--}$,
$b_{--}$, $\beta_{+++}$ ,$Y^{i}$ ,$\tau_{+}$, $\rho_{++}$, $\pa_{+}\f^{i}$)
and their derivatives, i.e.
\eq\ba{cl}
 \VV = &  {\it formal{\ }power{\ }series{\ }in{\ }the{\ }variables}
              {\ }\Lp  \pa^{l}_{-} \pa^{m}_{+} \xi, {\ }
               \pa^{l}_{-} \pa^{m+1}_{+} \f^{i}\Rp   \ , \\
       & \xi = \Lp h_{--}, B_{---}, c_{-}, \g_{--},
                   b_{--}, \beta_{+++} ,Y^{i} ,\tau_{+}, \rho_{++} \Rp
       \ , \qquad l,m=0,1,....
\ea\eqn{vv-space}
The solution of the descent equations \equ{desc-eq} will be
required to belong to $\VV$, i.e. $(\AA_{2}^{1}, \AA_{1}^{2}, \AA_{0}^{3})$
will be space-time forms whose coefficients are elements of the
local functional space \equ{vv-space}. We also remark that when
the integrated
consistency condition \equ{wz} is translated at the nonintegrated level
by means of the tower \equ{desc-eq} one is no more allowed to make integration
by parts. Therefore the fields and their derivatives have to be considered
as independent variables.

On the local space $\VV$ we introduce the two operators $\d_{+}, \d_{-}$
defined as
\eq
\d_{+} =  \pad{\ }{c_{-}} \ ,
\eqn{delta+}
and
\eq\ba{cl}
\d_{-} = \dsum{p=0}{\infty}\, \dsum{q=0}{\infty} \LP &
               \pa_{-}^{p}\,\pa_{+}^{q} h_{--}
              \pad{\ }{\pa_{-}^{p}\,\pa_{+}^{q} c_{-}} +
               \pa_{-}^{p}\,\pa_{+}^{q} B_{---}
              \pad{\ }{\pa_{-}^{p}\,\pa_{+}^{q} \g_{--}} -
                \pa_{-}^{p}\,\pa_{+}^{q} \tau_{+}
              \pad{\ }{\pa_{-}^{p}\,\pa_{+}^{q} b_{++}}  \\[2mm]
       & -      \pa_{-}^{p}\,\pa_{+}^{q} \rho_{++}
              \pad{\ }{\pa_{-}^{p}\,\pa_{+}^{q} \beta_{+++}} -
                   \pa_{-}^{p}\,\pa_{+}^{q} Y^{i}
              \pad{\ }{\pa_{-}^{p}\,\pa_{+}^{q+1} \f^{i}} \RP \ .
\ea\eqn{delta-}
They are easily seen to verify the algebraic relations
\eq
 \lac \BB_{\S},\d_{+} \rac = \pa_{+} \ , \qquad
 \lac \BB_{\S},\d_{-} \rac = \pa_{-} \ ,
\eqn{alg1}
and
\eq
 \lac \d_{-},\d_{+} \rac  = 0 \ , \qquad \lac \d_{\pm}, d \rac = 0 \ .
\eqn{alg2}
The operator $\d$,
\eq
   \d = dx^{+} \d_{+} + dx^{-} \d_{-} \ ,
\eqn{delta-op}
obeys thus
\eq
     d = - \lc \BB_{\S}, \d \rc  \ , \qquad  \lc \d, d \rc = 0 \ ,
\eqn{alg-delta-op}
i.e., according to \equ{dec},
the exterior space-time derivative $d$ has been decomposed
as a $BRS$ commutator.

Suppose now that the solution $\AA_{0}^{3}$ of the last equation
of the tower \equ{desc-eq} has been found, it is apparent then to check
that the higher cocycles
$\AA_{2}^{1}, \AA_{1}^{2}$ can be identified with the $\d$-transform of
$\AA_{0}^{3}$~\cite{silvio}:
\eq
\AA^{2}_{1} = \d \AA_{0}^{3} =
        dx^{+} \d_{+}\AA^{3}_{0} +  dx^{-} \d_{-}\AA^{3}_{0} \ ,
\eqn{aa21-sol}
and
\eq
 \AA^{1}_{2} = \frac{1}{2} \d^2  \AA^{3}_{0} =
        -dx^{+} dx^{-} \d_{+}  \d_{-}\AA^{3}_{0} \ .
\eqn{aa12-sol}
Let us proceed thus with the characterization of $\AA^{3}_{0}$, i.e.
with the computation of the cohomology of $\BB_{\S}$ in the sector of
the zero forms with ghost number and dimension three.

\section{Cohomology of $\BB_{\S}$ in the sector
of the zero  forms  with ghost number three and dimension three}

In order to study the cohomology of $\BB_{\S}$ we introduce
the filtering operator~\footnote{We recall here
that on the local space $\VV$ the action of the generic
functional differential
operator $\int {\OO} (\f)\fud{\ }{\f}$ is given by
\eq
\xint {\OO} (\f)\fud{\ }{ \f} :=
\dsum{p=0}{\infty} \dsum{q=0}{\infty}
 \pa_{-}^{p} \pa_{+}^{q} {\OO} (\f)\pad{\ }
                       {\pa_{-}^{p}\pa_{+}^{q}\f^{i}} \ .
\eqn{funct-act}
}
\eq
{\wti \NN} = \xint \Lp \f^{i} \fud{\ }{\f^{i}} +
           Y^{i} \fud{\ }{Y^{i}} \Rp \ ,
\eqn{ntilde-filt}
according to which the linearized operator $\BB_{\S}$ decomposes as
\eq
\BB_{\S}  = \BB^{(0)} +  \BB^{(1)} +  \BB^{(2)} +\BB^{(3)} \ ,
\eqn{first-decomp}
with
\eq
      \lc {\wti \NN},\BB^{(n)} \rc = n \BB^{(n)} \ ,
      \qquad n=0,1,2,3 \ .
\eqn{first-eigenv}
The operators $\BB^{(n)}$ in eq. \equ{first-decomp} are given in
{\bf {\it App.A}}. In particular the nilpotency of $\BB_{\S}$ implies that
\eq
  \dsum{j=0}{n} \BB^{(n-j)} \BB^{(j)} = 0 \ , \qquad  n=0,1,2,3
\eqn{first-nilp}
so that the operator $\BB^{(0)}$ turns out to be nilpotent as well
\eq
   \BB^{(0)} \BB^{(0)} =  0 \ .
\eqn{bb0-nilp}
The usefulness of the above decomposition relies on a very general
theorem on the $BRS$ cohomology~\cite{dixon,b3cd}. The latter states
that the cohomology of the full linearized Slavnov-Taylor
operator $\BB_{\S}$ is
isomorphic to a subspace of the cohomology of $\BB^{(0)}$.
Let us focus then on the study of the cohomology of the operator $\BB^{(0)}$.

\subsection{Cohomology of $\BB^{(0)}$ in the sector of the
 zero forms with ghost number three and dimension three}

In order to solve the equation
\eq
 \BB^{(0)} \Omega = 0 \ ,
\eqn{bb0-cohom}
with $\Omega$ a zero form with ghost number and dimension three,
we introduce a second filtering operator
\eq
   \NN= \xint \Lp
     c_{-} \fud{\ }{c_{-}} \, + \,
     h_{--} \fud{\ }{h_{--}}   \, + \,
     B_{---} \fud{\ }{B_{---}}  \, + \,
     \g_{--} \fud{\ }{\g_{--}} \Rp \ .
\eqn{second-filt}
It decomposes $\BB^{(0)}$ and $\Omega$ as
\eq
\BB^{(0)} = \d^{(0)} + \d^{(1)} +\d^{(2)} \ ,
\eqn{bb0-decomp}
and
\eq
  \Omega = \dsum{\n=0}{\infty} \Omega^{(\n)} \ ,
\eqn{omega-dec}
with
\eq
      \lc \NN ,\d^{(\n)} \rc = \n \d^{(\n)} \ ,  \qquad
      \NN \Omega^{(\n)} = \n \Omega^{(\n)} \ .
\eqn{second-eigenv}
Condition \equ{bb0-cohom} splits now into a system of equations
\eq
   \dsum{p=0}{\n} \d^{(p)} \Omega^{(\n-p)} = 0 \ , \qquad \n=0,1,2,3,...
\eqn{system}
As before, the nilpotency of $\BB^{(0)}$ implies
\eq
  \dsum{k=0}{q} \d^{(q-k)} \d^{(k)} = 0 \ , \qquad  q=0,1,2
\eqn{second-nilp}
and
\eq
  \d^{(0)} \d^{(0)} = 0 \ ,
\eqn{delta0-nilp}
so that, according to the previous theorem, the cohomology of $\BB^{(0)}$
is, in turn, isomorphic to a subspace of the cohomology of $\d^{(0)}$.
As we shall
see the operator $\d^{(0)}$ turns out to have a very elementary structure,
making the
computation of its cohomology rather easy. This will allow us to come
back to the operators $\BB^{(0)}$ and $\BB_{\S}$ and to identify their
cohomologies in a simple way.

Using {\bf {\it App.A}}, the operators $\d^{(0)}, \d^{(1)}, \d^{(2)}$ are
computed to be
\eq\ba{cl}
\d^{(0)}= \xint \Lp  & \pa_{-}c_{-} \fud{\ }{h_{--}}{\ } + {\ }
      \pa_{-}\g_{--}  \fud{\ }{B_{---}} {\ }- {\ }
      \pa_{+}\pa_{-}\f^{i}  \fud{\ }{Y^{i}} \\
   &  - {\ }\pa_{-}b_{++} \fud{\ }{\tau_{+}} {\ } - {\ }
      \pa_{-}\b_{+++} \fud{\ }{\r_{++}} \Rp  \ ,
\ea\eqn{delta0-exp}

\eq\ba{cl}
  \d^{(1)} = \xint  \LP &
      c_{-}\pa_{+}\f^{i}\fud{\ }{\f^{i}}{\ }
    + {\ }c_{-}\pa_{+}c_{-}\fud{\ }{c_{-}}
    + \Lp   c_{-}\pa_{+}h_{--} - h_{--}\pa_{+}c_{-} \Rp
     \fud{\ }{h_{--}} \\[2mm]
  & + \Lp c_{-}\pa_{+}\g_{--} + 2 \g_{--} \pa_{+} c_{-} \Rp
     \fud{\ }{\g_{--}}  \\[2mm]
  & +\Lp  2 \g_{--} \pa_{+}h_{--} - h_{--}\pa_{+} \g_{--}
    - 2 B_{---}\pa_{+} c_{-} +  c_{-} \pa_{+} B_{---} \Rp
      \fud{\ }{B_{---}} \\[2mm]
  & +\Lp c_{-}\pa_{+}b_{++} - 2 b_{++} \pa_{+} c_{-}
        + 2 \g_{--} \pa_{+} \b_{+++} - 3 \b_{+++} \pa_{+}\g_{--} \Rp
    \fud{\ }{b_{++}} \\[2mm]
  & + \Lp  c_{-}  \pa_{+}\b_{+++} -3  \b_{+++} \pa_{+} c_{-} \Rp
     \fud{\ }{\b_{+++}} \\[2mm]
  & +\Lp \pa_{+} h_{--} \pa_{+}\f^{i} +  h_{--}{\pa_{+}}^{2}\f^{i}
        - \pa_{+}(Y^{i} c_{-}) \Rp
      \fud{\ }{Y^{i}} \\[2mm]
  & + \Lp  h_{--} \pa_{+} \b_{+++} + 3 \b_{+++} \pa_{+} h_{--}
       + \pa_{+}(\r_{++} c_{-}) + 2 \r_{++} \pa_{+} c_{-} \Rp
    \fud{\ }{\r_{++}}  \\[2mm]
  & + \Lp 2 b_{++} \pa_{+} h_{--} + h_{--} \pa_{+} b_{++} +
        2 B_{---}\pa_{+}\b_{+++} + 3 \b_{+++}\pa_{+} B_{---} \Rp
    \fud{\ }{\tau_{+}} \\[2mm]
  & + \Lp 2 \tau_{+}\pa_{+} c_{-} +  c_{-} \pa_{+}\tau_{+}
     + 3 \r_{++}\pa_{+}\g_{--} + 2 \g_{--}\pa_{+} \r_{++} \Rp
     \fud{\ }{\tau_{+}}
   {\ }{\ }{\ }\RP  \ ,
\ea\eqn{delta1-op}
and
\eq\ba{cl}
 \d^{(2)} = \xint  \LP &
     \Lp 2  b_{++}\g_{--}\pa_{+}\g_{--}\pa_{+}\f^{i} \Rp
        \fud{\ }{\f^{i}} \\[2mm]
 & + \Lp 2 \pa_{+}( b_{++}\pa_{+} B_{---}\g_{--}\pa_{+}\f^{i} )
        -2 \pa_{+}( b_{++} B_{---}\pa_{+}\g_{--}\pa_{+}\f^{i} ) \Rp
        \fud{\ }{Y^{i}} \\[2mm]
 & - \Lp 2 \pa_{+}( Y^{i} b_{++}\g_{--}\pa_{+}\g_{--} )
       + 2 \pa_{+}(\tau_{+}\g_{--}\pa_{+}\g_{--}\pa_{+}\f^{i} ) \Rp
      \fud{\ }{Y^{i}} {\ }{\ }{\ }\RP
\ea\eqn{delta2-op}
Expression \equ{delta0-exp} shows that the variables
$(h, \pa_{-} c)$, $(B, \pa_{-}\g)$,
$(-Y^{i}, \pa_{+}\pa_{-}\f^{i})$, $(-\tau, \pa_{-} b)$, $(-\r, \pa_{-}\b)$
and their derivatives are grouped in $BRS$
doublets~\cite{poincare,dixon,b3cd}, i.e
\eq
     \d^{(0)}  \pa^{l}_{-} \pa^{m}_{+} u =  \pa^{l}_{-} \pa^{m}_{+} v  \ ,
     \qquad \d^{(0)}  \pa^{l}_{-} \pa^{m}_{+} v= 0 \ ,
\eqn{brs-doublets}
with
\eq\ba{l}
    u = \Lp h_{--}\ ,B_{---}\ ,-Y^{i}\ ,-\tau_{+}\ ,-\r_{++} \Rp \ , \es
    v = \Lp \pa_{-} c_{-} \ , \pa_{-}\g_{--} \ , \pa_{+}\pa_{-}\f^{i} \ ,
        \pa_{-}b_{++} \ , \pa_{-}\b_{+++} \Rp \ .
\ea\eqn{doubl-struct}
As it is well known, the cohomology does not depend on such
variables~\cite{dixon,b3cd}. It follows thus that on the local
space $\VV$ the cohomology
$\HH(\d^{(0)})$ of $\d^{(0)}$ is spanned by
\eq\ba{cl}
 \HH(\d^{(0)}) = &  {\it formal{\ }power{\ }series{\ }in{\ }the{\ }variables}
              {\ } \Lp  \pa^{i}_{+} \lambda, {\ }
                   \pa^{i+1}_{+} \f^{i} \Rp   \ , \\
         & \lambda = \Lp  c_{-}\ ,\g_{--}\ ,b_{++} \ ,\b_{+++} \Rp \ ,
           \qquad  i = 0,1,2,....
\ea\eqn{d0-cohom}
For the usefulness of the reader we display in {\bf {\it App.B}} all the
elements
of $\HH(\d^{(0)})$ belonging to the
sector of the zero forms with ghost number three and bounded
by dimension three.
The list is done according to the eigenvalues $\n$ of the filtering operator
$\NN$ of \equ{second-filt}. He will see, in particular, that the cohomology
spaces of $\d^{(0)}$ with $\n \ge 5$ are empty. This is easily seen to
imply that the system
\equ{system} reduces to a finite number of equations, i.e.
\eq\ba{l}
 \d^{(0)} \, \O^{(6)} \, + \, \d^{(1)} \, \O^{(5)} \, +
          \, \d^{(2)} \, \O^{(4)} \, = 0 \ ,  \es
 \d^{(0)} \, \O^{(5)} \, + \, \d^{(1)} \, \O^{(4)} \, +
          \, \d^{(2)} \, \O^{(3)} \, = 0 \ , \es
 \d^{(0)} \, \O^{(4)} \, + \, \d^{(1)} \, \O^{(3)} \, +
          \, \d^{(2)} \, \O^{(2)} \, = 0 \ , \es
 \d^{(0)} \, \O^{(3)} \, + \, \d^{(1)} \, \O^{(2)} \, +
          \, \d^{(2)} \, \O^{(1)} \, = 0  \ , \es
 \d^{(0)} \, \O^{(2)} \, + \, \d^{(1)} \, \O^{(1)} \, +
          \, \d^{(2)} \, \O^{(0)} \, = 0 \ , \es
 \d^{(0)} \, \O^{(1)} \, + \, \d^{(1)} \, \O^{(0)} \, = 0 \ , \es
 \d^{(0)} \, \O^{(0)} \, = 0 \ ,
\ea\eqn{finite-system}
the conditions with higher eingenvalues $(\n > 6)$ corresponding to a
trivial $\BB^{(0)}-$cocycle.

It is not difficult now, using {\bf {\it App.B}}, to work out the
restrictions imposed by the system \equ{finite-system} on the cohomology
$\HH(\d^{(0)})$ of $\d^{(0)}$. We shall not enter here into the technical
details of the computations, limiting ourselves to give the final result
concerning the operator $\BB^{(0)}$. It turns out that on the local
space $\VV$ the cohomology $\HH(\BB^{(0)})$ of $\BB^{(0)}$ in the sector
of zero forms with ghost number and dimension three contains five
elements, a representative of which may be choosen as
\eq\ba{l}
  \LP \Lp c_{-}\g_{--}\pa_{+}\g_{--}
     ( \pa_{+}\f^{i}\pa^{3}_{+}\f^{i} - \pa^{2}_{+}\f^{i}\pa^{2}_{+}\f^{i} )
    + T_{++} \pa_{+}\g_{--}
   ( \pa^{2}_{+} c_{-}\g_{--}
     + \dfrac{1}{2} c_{-}\pa^{2}_{+}\g_{--} ) \Rp, \es
{\ }{\ }{\ }    c_{-}\pa_{+}c_{-}\pa_{+}^{2}c_{-},{\ }
                          c_{-}\pa_{+}c_{-}\g_{--} W_{+++},{\ }
    c_{-}\g_{--}\pa_{+}\g_{--}T_{++}T_{++},{\ }
 c_{-}\pa_{+}c_{-}\g_{--}\pa_{+}\g_{--} b_{++} T_{++} \Rp
\ea\eqn{bb0-cohomolgy}
$W_{+++}$ and $T_{++}$ denoting the operators of equation \equ{t-w}.
\subsection{Completion of the cohomology of $\BB_{\S}$: a Russian formula}

Having characterized the cohomology of $\BB^{(0)}$ we turn to the
study of the cohomology of the complete linearized Slavnov-Taylor
operator $\BB_{\S}$
\equ{lin-slav}. Let us begin by observing that the cohomology
space \equ{bb0-cohomolgy} decomposes according to the eigenvalues of
the filtering operator ${\wti \NN}$ of eq.\equ{ntilde-filt} into the
subspaces
$\QQ^{(0)}, \QQ^{(2)}, \QQ^{(3)}, \QQ^{(4)}$
\eq
    {\wti \NN} \QQ^{(k)} = k  \QQ^{(k)} \ ,
\eqn{mm-decomp}
with
\eq
  \QQ^{(0)}  :=   c^{3} =  c_{-}\pa_{+} c_{-}\pa_{+}^{2} c_{-}  \ ,
\eqn{c3}
and
\eq
  \QQ^{(2)} =  m \QQ^{(2)}_{1} + p \QQ^{(2)}_{2} \ ,
\eqn{qq2}
$\QQ^{(2)}_{1}$, $\QQ^{(2)}_{2}$ given respectively by
\eq
   \QQ^{(2)}_{1} =
      c_{-}\pa_{+}c_{-}\g_{--}\pa_{+}\g_{--} b_{++} T_{++}  \ ,
\eqn{qq21}
\eq
  \QQ^{(2)}_{2} =
    c_{-}\g_{--}\pa_{+}\g_{--}
   \Lp \pa_{+}\f^{i}\pa^{3}_{+}\f^{i} - \pa^{2}_{+}\f^{i}\pa^{2}_{+}\f^{i} \Rp
  + T_{++} \pa_{+}\g_{--}
      \Lp  \pa^{2}_{+} c_{-}\g_{--}
           + \dfrac{1}{2} c_{-}\pa^{2}_{+}\g_{--} \Rp \ ,
\eqn{qq22}
$(m,p)$ being arbitrary coefficients. Finally, for $\QQ^{(3)}, \QQ^{(4)}$
we get
\eq
\QQ^{(3)} = c_{-}\pa_{+}c_{-}\g_{--}W_{+++}  \ ,
\eqn{qq3}
\eq
\QQ^{(4)} = c_{-}\g_{--}\pa_{+}\g_{--}T_{++}T_{++} \ .
\eqn{qq4}
In particular equations \equ{c3}-\equ{qq4} show that the cohomology of
$\BB_{\S}$ in the sector of zero forms with ghost number and dimension three
can contain at most five elements. Let us try then to see if the above
$\BB^{(0)}-$cocycles can be promoted to nontrivial cocycles of
$\BB_{\S}$, i.e. we look at the possibility of making a completion of
$(\QQ^{(0)}, \QQ^{(2)}_{1}, \QQ^{(2)}_{2}, \QQ^{(3)}, \QQ^{(4)})$ to
obtain elements of the cohomology of $\BB_{\S}$. This means find,
for each $\QQ^{(i)}$, a term $\RR_{\QQ^{(i)}}$
\eq
   \QQ^{(i)} \rightarrow {\hat \QQ^{(i)}} = \QQ^{(i)} + \RR_{\QQ^{(i)}} \ ,
\eqn{rr-term}
such that
\eq\ba{cl}
  & i) \qquad  \BB_{\S}  {\hat \QQ^{(i)}} = 0 \ , \\[4mm]
  & ii) \qquad {\hat \QQ^{(i)}}  \ne \BB_{\S} - variation  \ .
\ea\eqn{completion}
Of course the $\RR_{\QQ^{(i)}}$'s are always defined modulo trivial
$\BB_{\S}-$cocycles.

The problem of the completion of the cohomology of $\BB_{\S}$ is in
general rather complicated and, depending on the Lie algebra
structure of the generators of the BRS symmetry,
requires the knowledge of the corresponding Lie algebra cohomolgy
(see for instance refs.~\cite{gauge} for the case of Yang-Mills).
However we shall see that in the present case it is greatly
simplified thanks to the existence of a {\it Russian-like}
formula~\cite{russian} which allows to map
elements of the cohomology of $\BB^{(0)}$ into nontrivial
cocycles of $\BB_{\S}$ in a very simple and elegant way.

Let us proceed thus by discussing in detail the completion of the cocycle
$c^3$ of equation \equ{c3}. To this purpose we introduce the following
combination, here called the {\it Russian-like} formula,
\eq
     {\hat c_{-}} = c_{-} - 2 b_{++}\g_{--}\pa_{+}\g_{--} \ .
\eqn{chat-variable}
Using {\bf {\it App.A}} one easily verifies that ${\hat c_{-}}$ transforms
under $\BB_{\S}$ in the same way as $c_{-}$ transforms under
$\BB^{(0)}$, i.e.
\eq\ba{l}
   \BB_{\S} {\hat c_{-}} = {\hat c_{-}}{\pa_{+}}{\hat c_{-}} \ , \es
   \BB^{(0)}  c_{-} =  c_{-} \pa_{+}  c_{-} \ .
\ea\eqn{hatc-c-transf}
It is apparent then from equation \equ{bb0-cohomolgy} that the cocycle
\eq
{\hat c}^{3} = {\hat c_{-}}\pa_{+}{\hat c_{-}}\pa_{+}^{2}{\hat c_{-}}
\eqn{hat-c3}
is $\BB_{\S}-$invariant. It satisfies thus the first requirement
of \equ{completion}.

\noindent The expression \equ{hat-c3} is computed to be
\eq
 {\hat c}^{3} = c^{3} + \RR_{c^{3}}   \ ,
\eqn{express-hatc3}
with
\eq\ba{cl}
 \RR_{c^{3}} =
        & -2 c_{-}\pa_{+}c_{-} (\pa_{+}^{2}b_{++}\g_{--}\pa_{+}\g_{--} +
           2 \pa_{+}b_{++}\g_{--}\pa_{+}^{2}\g_{--} ) \\
        & -2 c_{-}\pa_{+}c_{-} ( b_{++}\pa_{+}\g_{--}\pa^{2}\g_{--} +
             b_{++}\g_{--}\pa_{+}^{3}\g_{--} )  \\
       & + 2 c_{-}\pa_{+}^{2}c_{-} (\pa_{+}b_{++}\g_{--}\pa_{+}\g_{--} +
           b_{++}\g_{--}\pa_{+}^{2}\g_{--} ) \\
       &  - 2 \pa_{+}c_{-}\pa_{+}^{2}c_{-}b_{++}\g_{--}\pa_{+}\g_{--} \ .
\ea\eqn{rr-express}
Notice that $\RR_{c^{3}}$ belongs to the subspace of the
filtering operator ${\wti \NN}$ with zero eigenvalue. In addition it
turns out to be an exact $\BB^{(0)}-$cocycle
\eq\ba{cl}
 \RR_{c^{3}}  & =   \BB^{(0)} \MM^{(0)} \\
  & = \BB^{(0)}  \Lp  4 \pa_{+}^{2} c_{-} \g_{--} \pa_{+}
        \g_{--} b_{++}  -
      2 \pa_{+} c_{-} \g_{--}  \pa_{+}^{2} \g_{--} b_{++} \\[2mm]
  & {\ }{\ }{\ }{\ }{\ }{\ }{\ }{\ }{\ }{\ }
    - 2 \pa_{+} c_{-} \g_{--}  \pa_{+} \g_{--} \pa_{+} b_{++} +
      \dfrac{3}{4}  c_{-} \pa_{+} \g_{--}  \pa_{+}^{2} \g_{--}b_{++}  \\[2mm]
  & {\ }{\ }{\ }{\ }{\ }{\ }{\ }{\ }{\ }{\ }
   + \dfrac{3}{2} \g_{--} \pa_{+} \g_{--} \pa_{+}^{2} \g_{--}\b_{+++}
   - \dfrac{1}{2} c_{-} \g_{--}  \pa_{+}^{3} \g_{--}b_{++}  \Rp \ .
\ea\eqn{rr-triv}
Let us now prove that \equ{express-hatc3} also satisfies the second
requirement of \equ{completion}. As usual~\cite{b3cd} we proceed by
assuming the converse. Suppose then that
\eq
    {\hat c}^{3} =  \BB_{\S} \Xi \ ,
\eqn{c3-triv}
for some local formal power series $\Xi$.
Decomposing \equ{c3-triv} according to the eingenvalues
of ${\wti \NN}$ and taking into account that both $c^3$ and $\RR_{c^{3}}$
belong to the subspace with zero eigenvalue, one gets
\eq
  c^{3} + \RR_{c^{3}} = \BB^{(0)} \Xi^{(0)} \ .
\eqn{decomp-c3}
This condition, due to the $\BB^{(0)}-$triviality of $\RR_{c^{3}}$
(eq. \equ{rr-triv}), yields
\eq
 c^{3} = \BB^{(0)} ( \Xi ^{(0)} - \MM^{(0)}) \ ,
\eqn{contradict}
which would imply that $c^{3}$ is $\BB^{(0)}-$trivial, in contradiction
with \equ{bb0-cohomolgy}. We have thus proven that expression \equ{hat-c3}
defines
a cohomology class of the operator $\BB_{\S}$ in the sector of zero forms
with ghost number and dimension three.
In particular, equation \equ{chat-variable} has been shown to give
a simple and elegant
way of obtaining cohomology classes of $\BB_{\S}$ from those corresponding
to $\BB^{(0)}$, justifying then the name of {\it Russian-like} formula.

Let us proceed to present the completion of the remaining cocycles
\equ{qq2}-\equ{qq4}. Concerning expression $\QQ^{(4)}$ of
eq. \equ{qq4} one easily checks that it is $\BB_{\S}-$invariant
\eq
\BB_{\S}  \QQ^{(4)} = 0 \ .
\eqn{invariant-qq4}
However, due to
\eq
\QQ^{(4)} = - \BB_{\S} (c_{-}\g_{--}\pa_{+}\g_{--}b_{++}T_{++}) \ ,
\eqn{triv-qq4}
it is cohomologically trivial. Let us consider then the
cocycle $\QQ^{(2)}_{2}$ in \equ{qq22}. Using the {\it Russian-like}
formula \equ{chat-variable}
it is not difficult to prove that a completion of $\QQ^{(2)}_{2}$
is given by
\eq
    \QQ^{(2)}_{2} \rightarrow {\hat  \QQ^{(2)}_{2}} = \QQ^{(2)}_{2}
 + \RR_{\QQ^{(2)}_{2}} \ ,
\eqn{qq22-complet}
with $\RR_{\QQ^{(2)}_{2}}$
\eq
 \RR_{\QQ^{(2)}_{2}} = 3 {\hat c}_{-} \g_{--} \pa_{+} \g_{--}
       \Lp   \pa_{+}^{3} \g_{--}  \b_{+++} +
            \pa_{+}^{2} \g_{--} \pa_{+} \b_{+++} \Rp \ .
\eqn{rr-qq22-express}
As in the case of the cocycle ${\hat c}^3$, expression
\equ{qq22-complet} can be proven to be $\BB_{\S}-$nontrivial.
It identifies thus a cohomology class.

Finally, the two cocycles $\QQ^{(2)}_{1}$ and $\QQ^{(3)}$
yield a $\BB_{\S}-$invariant quantity only in the
combination
\eq
 \QQ^{(3)} + 4 \QQ^{(2)}_{1}  \ ,
\eqn{comb}
wich is seen to be trivial, i.e.
\eq
 \Lp \QQ^{(3)} + 4 \QQ^{(2)}_{1} \Rp = - \BB_{\S}
            ( c_{-}\pa_{+} c_{-} \g_{--} \b_{+++} ) \ .
\eqn{triv-comb}
This concludes the analysis of the completion of the cohomology of
$\BB_{\S}$.

\subsection{Summary}

Let us summarize here, for the convenience of the reader, the final result
on the cohomology of $\BB_{\S}$.
On the local space $\VV$ the cohomology of the linearized operator $\BB_{\S}$
in the sector of zero forms $\AA_{0}^{3}$
(see eq. \equ{desc-eq}) with ghost number and dimension three contains only
two elements. They can be written as
\eq
  \AA_{0}^{3} = \Lp {\hat \QQ^{(0)}}, {\ } {\hat  \QQ^{(2)}_{2}} \Rp
\eqn{aa03-express}
with
\eq\ba{cl}
 {\hat \QQ^{(0)}} =  {\hat c^3}  = &
         c_{-}\pa_{+} c_{-}\pa_{+}^{2} c_{-}  \\
      & -2 c_{-}\pa_{+}c_{-} (\pa_{+}^{2}b_{++}\g_{--}\pa_{+}\g_{--} +
           2 \pa_{+}b_{++}\g_{--}\pa_{+}^{2}\g_{--} ) \\
      & -2 c_{-}\pa_{+}c_{-} ( b_{++}\pa_{+}\g_{--}\pa^{2}\g_{--} +
             b_{++}\g_{--}\pa_{+}^{3}\g_{--} )  \\
      & + 2 c_{-}\pa_{+}^{2}c_{-} (\pa_{+}b_{++}\g_{--}\pa_{+}\g_{--} +
           b_{++}\g_{--}\pa_{+}^{2}\g_{--} ) \\
      &  - 2 \pa_{+}c_{-}\pa_{+}^{2}c_{-}b_{++}\g_{--}\pa_{+}\g_{--} \ ,
\ea\eqn{aao0-qqo}
and
\eq\ba{cl}
 {\hat  \QQ^{(2)}_{2}} = &
        {\hat c_{-}}\g_{--}\pa_{+}\g_{--}
   \Lp \pa_{+}\f^{i}\pa^{3}_{+}\f^{i} - \pa^{2}_{+}\f^{i}\pa^{2}_{+}\f^{i} \Rp
       \\
 & +{\ } T_{++} \pa_{+}\g_{--}
   \Lp   \pa^{2}_{+} {\hat c_{-}}\g_{--}
           + \dfrac{1}{2} {\hat c_{-}}\pa^{2}_{+}\g_{--} \Rp \\
 &  + {\ } 3 {\hat c}_{-} \g_{--} \pa_{+} \g_{--}
       \Lp   \pa_{+}^{3} \g_{--}  \b_{+++} +
            \pa_{+}^{2} \g_{--} \pa_{+} \b_{+++} \Rp \ .
\ea\eqn{aao3-qq22}

\section{Anomalies}

Having characterized the cohomology of $\BB_{\S}$ in the sector of the
zero forms with ghost number and dimension three we are now ready to
solve the ladder \equ{desc-eq} and find the anomalies which occur in
the quantum extension of the Slavnov-Taylor identity \equ{slavnov}.

Using the decomposition \equ{alg-delta-op} and the equations
\equ{aa21-sol}, \equ{aa12-sol} it is immediate to check that to each
element of $\AA_{0}^{3}$ in \equ{aa03-express} corresponds a two form
$\AA_{2}^{1}$ with ghost number one which is a nontrivial solution of the
consistency condition \equ{loc-wz}
\eq
  \AA_{2}^{1} = \Lp \UU_{2}^{1}, {\ }\TT_{2}^{1} \Rp \ ,
\eqn{aa21-express}
with $\UU_{2}^{1}$ and $\TT_{2}^{1}$ given respectively by
\eq
  \UU_{2}^{1} =  -dx^{+} dx^{-} \d_{+}  \d_{-}  {\hat \QQ^{(0)}}, \qquad
  \TT_{2}^{1} =  -dx^{+} dx^{-} \d_{+}  \d_{-}  {\hat  \QQ^{(2)}_{2}} \ .
\eqn{uu-tt}
They read
\eq\ba{cl}
   \UU_{2}^{1} =  dx^{+} dx^{-} \LP
 &    \pa_{+}h_{--}\pa_{+}^{2}c_{-} -  \pa_{+}^{2}h_{--}\pa_{+}c_{-}
     -2 \pa_{+}h_{--}\pa_{+}^{2}b_{++}\g_{--}\pa_{+}\g_{--} \\[2mm]
 &   -2 \pa_{+}^{2}\tau_{+}\pa_{+}c_{-}\g_{--}\pa_{+}\g_{--}
     -2 B_{---}\pa_{+}c_{-}\pa_{+}^{2}b_{++}\pa_{+}\g_{--} \\[2mm]
 &   +2 \pa_{+}B_{---}\pa_{+}c_{-}\pa_{+}^{2}b_{++}\g_{--}   \\[2mm]
 &  -4 \pa_{+}h_{--}\pa_{+}b_{++}\g_{--}\pa^{2}\g_{--}
    -4 \pa_{+}\tau_{+}\pa_{+}c_{-}\g_{--}\pa_{+}^{2}\g_{--}   \\[2mm]
 &  -4 B_{---}\pa_{+}c_{-}\pa_{+}b_{++}\pa_{+}^{2}\g_{--}
    +4 \pa_{+}^{2}B_{---}\pa_{+}c_{-}\pa_{+}b_{++}\g_{--} \\[2mm]
 &  -2 \pa_{+}h_{--}b_{++}\pa_{+}\g_{--}\pa_{+}^{2}\g_{--}
    -2 \tau_{+}\pa_{+}c_{-}\pa_{+}\g_{--}\pa_{+}^{2}\g_{--}  \\[2mm]
 &  -2 \pa_{+}B_{---}\pa_{+}c_{-}b_{++}\pa_{+}^{2}\g_{--}
    +2\pa_{+}^{2}B_{---}\pa_{+}c_{-}b_{++}\pa_{+}\g_{--}  \\[2mm]
 &  -2 \pa_{+}h_{--}b_{++}\g_{--}\pa_{+}^{3}\g_{--}
    -2\tau_{+}\pa_{+}c_{-}\g_{--}\pa_{+}^{3}\g_{--}  \\[2mm]
 &  -2 B_{---}\pa_{+}c_{-}b_{++}\pa_{+}^{3}\g_{--}
    +2\pa_{+}^{3}B_{---}\pa_{+}c_{-}b_{++}\g_{--}  \\[2mm]
 &  +2 \pa_{+}^{2}h_{--}\pa_{+}b_{++}\g_{--}\pa_{+}\g_{--}
    +2 \pa_{+}\tau_{+}\pa_{+}^{2}c_{-}\g_{--}\pa_{+}\g_{--}  \\[2mm]
 &  +2 B_{---}\pa_{+}^{2}c_{-}\pa_{+}b_{++}\pa_{+}\g_{--}
    -2 \pa_{+}B_{---}\pa_{+}^{2}c_{-}\pa_{+}b_{++}\g_{--}  \\[2mm]
 &  +2 \pa_{+}^{2}h_{--}b_{++}\g_{--}\pa_{+}^{2}\g_{--}
    +2 \tau_{+}\pa_{+}^{2}c_{-}\g_{--}\pa_{+}^{2}\g_{--}  \\[2mm]
 &  +2 B_{---}\pa_{+}^{2}c_{-}b_{++}\pa_{+}^{2}\g_{--}
    -2 \pa_{+}^{2}B_{---}\pa_{+}^{2}c_{-}b_{++}\g_{--} \,  \RP \ ,
\ea\eqn{uu21-express}
and
\eq\ba{cl}
  \TT_{2}^{1} =  dx^{+} dx^{-} \LP
 &  B_{---} \pa_{+} \g_{--}\pa_{+}^{3} \f^{i} \pa_{+} \f^{i} \, - \,
    \pa_{+} B_{---} \g_{--}\pa_{+}^{3} \f^{i} \pa_{+} \f^{i} \\[2mm]
 &  - \g_{--} \pa_{+} \g_{--}\pa_{+}^{2} Y^{i} \pa_{+} \f^{i} \, - \,
    \g_{--} \pa_{+} \g_{--} Y^{i}\pa_{+}^{3} \f^{i}   \\[2mm]
 & - B_{---} \pa_{+} \g_{--}\pa_{+}^{2} \f^{i} \pa_{+}^{2} \f^{i} \, + \,
   \g_{--} \pa_{+} B_{---}\pa_{+}^{2} \f^{i} \pa_{+}^{2} \f^{i}  \\[2mm]
 &  + 2  \g_{--} \pa_{+} \g_{--}\pa_{+} Y^{i} \pa_{+}^{2} \f^{i} \, - \,
   \dfrac{1}{2} \pa_{+} B_{---} \pa_{+}^{2} \g_{--} T_{++}  \\[2mm]
 & + \dfrac{1}{2} \pa_{+}^{2} B_{---} \pa_{+} \g_{--} T_{++} \, +  \,
   \dfrac{1}{2} \pa_{+} \g_{--} \pa_{+}^{2} \g_{--} Y^{i} \pa_{+} \f^{i}
      \\[2mm]
 & + 3 B_{---} \pa_{+} \g_{--} \pa_{+}^{3} \g_{--} \b_{+++} \,
   - \, 3 \pa_{+} B_{---}  \g_{--} \pa_{+}^{3} \g_{--} \b_{+++} \\[2mm]
 & + 3 \pa_{+}^{3} B_{---} \g_{--} \pa_{+} \g_{--} \b_{+++} \,
   +  3 \g_{--} \pa_{+} \g_{--} \pa_{+}^{3} \g_{--} \r_{++}  \\[2mm]
 & +  3 B_{---} \pa_{+} \g_{--} \pa_{+}^{2} \g_{--} \pa_{+} \b_{+++} \,
   -  3 \pa_{+} B_{---} \g_{--} \pa_{+}^{2} \g_{--} \pa_{+} \b_{+++} \\[2mm]
 & +  3 \pa_{+}^{2} B_{---} \g_{--} \pa_{+} \g_{--} \pa_{+} \b_{+++} \,
   +  3 \g_{--} \pa_{+} \g_{--} \pa_{+}^{2} \g_{--} \pa_{+} \r_{++}  \RP \ .
\ea\eqn{tt21-express}
Let us remember that, in general~\cite{silvio},
eqs. \equ{aa21-sol}-\equ{aa12-sol}
give only a particular solution of the descent equations \equ{desc-eq}.
However, in the present case, \equ{uu21-express} and \equ{tt21-express}
yield the most general solution.
This is due to the fact that the cohomology of $\BB_{\S}$ turns out to be
vanishing in the sectors of the one forms with ghost number two and of the
two forms with ghost number one~\footnote{This property follows from the
emptiness of the cohomology of $\BB^{(0)}$ in the form sectors considered
above, as it is easily proven by repeating the same analysis
of the previous section.}.

Finally, for the integrated cocycles we get
\eq\ba{cl}
\AA_{\UU}  & = \dint \UU_{2}^{1}  \\
 &           = \xint \Lp
        2c_{-} \pa_{+}^{3}h_{--}
       - 4 \pa_{+}^{3}h_{--} \g_{--} \pa_{+}\g_{--} b_{++}
       + 4 \pa_{+}^{3} c_{-} B_{---} \pa_{+}\g_{--} b_{++}  \\[2mm]
 &  {\ }{\ }{\ }{\ }{\ }{\ }{\ }{\ }{\ }{\ }{\ }{\ }
        - 4 \pa_{+}^{3}c_{-} \g_{--} \pa_{+}B_{---}  b_{++}
       -  4 \pa_{+}^{3}c_{-}\g_{--}\pa_{+}\g_{--}\tau_{+} \Rp  \ ,
\ea\eqn{aauu-anomaly}
and
\eq\ba{cl}
\AA_{\TT} & = \dint\TT_{2}^{1} \\
 &          = \xint \LP {\ }
        2\Lp B_{---} \pa_{+} \g_{--}  -   \pa_{+} B_{---} \g_{--} \Rp
       \pa_{+}^{3} \f^{i} \pa_{+} \f^{i}  \\[2mm]
 &  {\ }{\ }{\ }{\ }{\ }{\ }{\ }{\ }{\ }{\ }{\ }{\ }
       +   \dfrac{3}{2}\Lp  \pa_{+}^{2} B_{---} \pa_{+} \g_{--}  -
            \pa_{+} B_{---} \pa_{+}^{2} \g_{--} \Rp  T_{++} \\[2mm]
 &  {\ }{\ }{\ }{\ }{\ }{\ }{\ }{\ }{\ }{\ }{\ }{\ }
       + \Lp \pa_{+}^{3} B_{---} \g_{--} -
        B_{---} \pa_{+}^{3} \g_{--} \Rp T_{++} \\[2mm]
 &  {\ }{\ }{\ }{\ }{\ }{\ }{\ }{\ }{\ }{\ }{\ }{\ }
      -  \g_{--} \pa_{+} \g_{--}\pa_{+}^{2} Y^{i} \pa_{+} \f^{i} -
        \g_{--} \pa_{+} \g_{--} Y^{i}  \pa_{+}^{3} \f^{i}  \\[2mm]
 &  {\ }{\ }{\ }{\ }{\ }{\ }{\ }{\ }{\ }{\ }{\ }{\ }
     +  2  \g_{--} \pa_{+} \g_{--}\pa_{+} Y^{i} \pa_{+}^{2} \f^{i}  +
       \dfrac{1}{2} \pa_{+} \g_{--} \pa_{+}^{2} \g_{--}
        Y^{i} \pa_{+} \f^{i}  \RP \ .
\ea\eqn{aatt-anomaly}
Equations \equ{aauu-anomaly}, \equ{aatt-anomaly} give the expressions of
the anomalies which arise in chiral $\WW_3-$gravity.
They are in complete agreement with the one loop result of the
Feynman graphs computation done in ref.~\cite{hl}.
In particular, the term $\AA_{\UU}$ of \equ{aauu-anomaly},
also called the universal gravitational anomaly, is easily seen to be a
generalization of the usual diffeomorphism anomaly of
the bosonic string~\cite{beltrami}, while the second term
$\AA_{\TT}$ in eq.\equ{aatt-anomaly} is a matter-dependent anomaly
whose origin can be traced back to the nonlinearity of the
$\WW_{3}-$algebra~\cite{hl,nonchiral}.

\section{Triviality of the classical action}

This section is devoted to display an interesting algebraic
property of the classical $\WW_3-$action of eq.\equ{tot-act}.

As it happens in the case of the bosonic string~\cite{wien} (see also
ref.~\cite{swien} for the generalization to superstring), the
classical action \equ{tot-act} turns out to be cohomologically trivial, i.e.
it is a pure $\BB_{\S}-$variation.
It is easily checked indeed that
\eq
\S = \BB_{\S} \xint \Lp \frac{1}{2} Y^{i}\f^{i}
  - \tau_{+} c_{-} - \frac{1}{2} \r_{++}\g_{--}
  -\frac{1}{2} \b_{+++} B_{---} \Rp \ .
\eqn{triv-action}
This property allows to interpret in a suggestive way the
$\WW_3-$gravity as a topological model of the Witten's type~\cite{topmodels}.

\vspace{2cm}

\noindent{\large{\bf Acknowledgements}}

The {\it Conselho Nacional de Desenvolvimento Cientifico e Tecnologico},
$CNPq$-Brazil is gratefully acknowledged for the financial support.

\newpage
\appendix

\section{The linearized Slavnov-Taylor operator $\BB_{\S}$}

As shown in Sect. 4, the linearized operator $\BB_{\S}$ decomposes
according to the eigenvalues of the filtering operator ${\wti \NN}$ of
eq. \equ{ntilde-filt} as
\eq
\BB_{\S}  = \BB^{(0)} +  \BB^{(1)} +  \BB^{(2)} +\BB^{(3)} \ .
\eqn{app-first-decomp}
with
\eq
      \lc {\wti \NN},\BB^{(n)} \rc = n \BB^{(n)} \ ,
      \qquad n=0,1,2,3 \ .
\eqn{first-eigenv}
The operators $\BB^{(n)}$ read explicitely
\eq\ba{cl}
\BB^{(0)} = \xint &\LP
     \Lp c_{-}\pa_{+}\f^{i}
     + 2  b_{++}\g_{--}\pa_{+}\g_{--}\pa_{+}\f^{i} \Rp
        \fud{\ }{\f^{i}} \\[2mm]
  &  + {\ }c_{-}\pa_{+}c_{-}\fud{\ }{c_{-}} {\ }
     + {\ } \Lp  \pa_{-}c_{-}  + c_{-}\pa_{+}h_{--} - h_{--}\pa_{+}c_{-} \Rp
        \fud{\ }{h_{--}} \\[2mm]
  &  + \Lp c_{-}\pa_{+}\g_{--} + 2 \g_{--} \pa_{+} c_{-} \Rp
     \fud{\ }{\g_{--}} \\[2mm]
  &  +\Lp - 2 B_{---}\pa_{+} c_{-} +  c_{-} \pa_{+} B_{---} \Rp
      \fud{\ }{B_{---}} \\[2mm]
  &  +\Lp \pa_{-}\g_{--}  + 2 \g_{--} \pa_{+}h_{--} -
      h_{--}\pa_{+} \g_{--}\Rp  \fud{\ }{B_{---}} \\[2mm]
  &  +\Lp c_{-}\pa_{+}b_{++} - 2 b_{++} \pa_{+} c_{-}
        + 2 \g_{--} \pa_{+} \b_{+++} - 3 \b_{+++} \pa_{+}\g_{--} \Rp
     \fud{\ }{b_{++}} \\[2mm]
  &  + \Lp  c_{-}  \pa_{+}\b_{+++} -3  \b_{+++} \pa_{+} c_{-} \Rp
      \fud{\ }{\b_{+++}} \\[2mm]
  &  +\Lp  - {\ } \pa_{+}\pa_{-}\f^{i}   + \pa_{+} h_{--} \pa_{+}\f^{i}
     +  h_{--}{\pa_{+}}^{2}\f^{i}
     + 2 \pa_{+}( b_{++}\pa_{+} B_{---}\g_{--}\pa_{+}\f^{i} ) \\
  &  {\ }{\ }{\ }{\ }{\ }
        -2 \pa_{+}( b_{++} B_{---}\pa_{+}\g_{--}\pa_{+}\f^{i} )
        - \pa_{+}(Y^{i} c_{-})
	 - 2 \pa_{+}( Y^{i} b_{++}\g_{--}\pa_{+}\g_{--} ) \\
  &  {\ }{\ }{\ }{\ }{\ }
     - 2 \pa_{+}(\tau_{+}\g_{--}\pa_{+}\g_{--}\pa_{+}\f^{i} )
           \Rp \fud{\ }{Y^{i}} \\[2mm]
  &  + \Lp  - {\ } \pa_{-}\b_{+++}
     + h_{--} \pa_{+} \b_{+++} + 3 \b_{+++} \pa_{+} h_{--} \\[2mm]
  &  {\ }{\ }{\ }{\ }{\ }
       + \pa_{+}(\r_{++} c_{-}) + 2 \r_{++} \pa_{+} c_{-} \Rp
    \fud{\ }{\r_{++}}  \\[2mm]
  & + \Lp  - {\ }\pa_{-}b_{++}  + 2 b_{++} \pa_{+} h_{--}
      + h_{--} \pa_{+} b_{++} +
        2 B_{---}\pa_{+}\b_{+++}   +  c_{-} \pa_{+}\tau_{+} \\
 &  {\ }{\ }{\ }{\ }{\ } + 3 \b_{+++}\pa_{+} B_{---}
 	 + 2 \tau_{+}\pa_{+} c_{-}
     + 3 \r_{++}\pa_{+}\g_{--} + 2 \g_{--}\pa_{+} \r_{++} \Rp
    \fud{\ }{\tau_{+}} {\ }\RP  \ ,
\ea\eqn{bb0-expression}
\eq\ba{cl}
\BB^{(1)}= \xint &\LP
  \lp   d_{ijk}\g_{--} \pa_{+} \f^{j} \pa_{+} \f^{k}
          \rp \fud{\ }{\f^{i}} \\[2mm]
& + {\ }\Lp   d_{ijk}\pa_{+} B_{---}
   \pa_{+} \f^{j} \pa_{+} \f^{k}
   + 2  d_{ijk} B_{---} \pa_{+} \f^{j} \pa_{+}^{2} \f^{k} \\
&  {\ }{\ }{\ }{\ }{\ } + 2  \pa_{+} (  d_{ijk} \g_{--} Y^{j}
  \pa_{+} \f^{k} ) \Rp
      \fud{\ }{Y^{i}}  {\ } {\ } {\ }\RP \ ,
\ea\eqn{bb1-expression}
\eq\ba{cl}
\BB^{(2)}= \xint &\LP
  \Lp  2 \g_{--} \pa_{+} \g_{--} T_{++} \Rp \fud{\ }{c_{-}} {\ }
           + T_{++}     \fud{\ }{b_{++}}
          - Y^{i}  \pa_{+} \f^{i}    \fud{\ }{\tau_{+}} \\[2mm]
  &{\ } + \Lp 2( \pa_{+} B_{---} \g_{--} - B_{---} \pa_{+} \g_{--})  T_{++}
           + 2 Y^{i} \g_{--} \pa_{+} \g_{--}  \pa_{+} \f^{i}
            \Rp     \fud{\ }{h_{--}} \\[2mm]
 &{\ } + \Lp -2 \pa_{+}  b_{++} \g_{--}  T_{++}
          -4 b_{++}  \pa_{+} \g_{--}  T_{++}
	  -2 b_{++} \g_{--} \pa_{+}  T_{++}
	  \Rp   \fud{\ }{\b_{+++}} \\[2mm]
 &{\ } + \Lp 4 b_{++} \pa_{+} B_{---} T_{++} +
          2  \pa_{+} b_{++}   B_{---} T_{++} +
          2 b_{++}  B_{---} \pa_{+} T_{++} \\[2mm]
&{\ }   {\ }{\ }{\ }{\ }{\ }
       +  4 Y^{i} b_{++}  \pa_{+} \g_{--}  \pa_{+} \f^{i} +
          2 \pa_{+} Y^{i} b_{++}   \g_{--}  \pa_{+} \f^{i} \\[2mm]
&{\ }  {\ }{\ }{\ }{\ }{\ } +
          2 Y^{i} \pa_{+} b_{++}  \g_{--}  \pa_{+} \f^{i}
       +  2 Y^{i} b_{++} \g_{--}  \pa_{+}^{2} \f^{i} +
	  4 \tau_{+}  \pa_{+} \g_{--}  T_{++} \\[2mm]
&{\ }  {\ }{\ }{\ }{\ }{\ } +
	  2 \pa_{+} \tau_{+} \g_{--} T_{++}
        + 2 \tau_{+} \g_{--}  \pa_{+}^{2} \f^{i}
               \pa_{+} \f^{i} \Rp \fud{\ }{\r_{++}}
             {\ } {\ } {\ }\RP \ ,
\ea\eqn{bb2-expression}
and
\eq
\BB^{(3)}= \xint \LP
   W_{+++} \fud{\ }{\b_{+++}} -  d_{ijk} Y^{i}
               \pa_{+} \f^{j} \pa_{+} \f^{k} \fud{\ }{\r_{++}}
	       {\ } {\ } {\ } \RP \ .
\eqn{bb3-expression}

\newpage
\section{Cohomology of $\d^{(0)}$}

We give here the list of all the elements belonging to the cohomology
of $\HH(\d^{(0)})$ of the operator $\d^{(0)}$ of eq. \equ{delta0-exp}
in the sector $\L$ of the zero forms
with gost number and dimension three. The list is done according
to the eigenvalues $\n$ of the filtering operator
$\NN$ of \equ{second-filt}.
It turns out that the only nonvanishing subspaces are those corresponding
to the eigenvalues $3$ and $4$ of $\NN$, i.e.
\eq
 \L = \L^{(3)} + \L^{(4)}  \ ,
\eqn{lambda-space}
with $\L^{(3)}$ and $\L^{(4)}$ given respectively by
\eq\ba{cl}
 \L^{(3)}  =  \Lp
 &
     \pa_{+}^{4} c_{-}  \g_{--}  \pa_{+} \g_{--}, {\ }{\ }{\ }
     \g_{--} \pa_{+} \g_{--} \pa_{+}^{5} \g_{--}, {\ }{\ }{\ }
     \g_{--} \pa_{+}^{2} \g_{--} \pa_{+}^{4} \g_{--}, \\[2mm]

 &     c_{-}\pa_{+} c_{-}\pa_{+}^{2} c_{-}, {\ }{\ }{\ }
       c_{-}\pa_{+}^{4} c_{-} \g_{--},  {\ }{\ }{\ }
       \pa_{+} c_{-}\pa_{+}^{3} c_{-} \g_{--},  \\[2mm]

 &     c_{-}\pa_{+}^{3} c_{-} \pa_{+} \g_{--}, {\ }{\ }{\ }
       \pa_{+} c_{-}\pa_{+}^{2} c_{-}  \pa_{+} \g_{--}, {\ }{\ }{\ }
       c_{-}\pa_{+}^{2} c_{-}  \pa_{+}^{2} \g_{--}, \\[2mm]

 &     c_{-}\pa_{+} c_{-}  \pa_{+}^{3} \g_{--}, {\ }{\ }{\ }
       c_{-} \g_{--}  \pa_{+}^{5} \g_{--}, {\ }{\ }{\ }
       c_{-}  \pa_{+} \g_{--}  \pa_{+}^{4} \g_{--},  \\[2mm]
 &
      c_{-}  \pa_{+}^{2} \g_{--} \pa_{+}^{3} \g_{--}, {\ }{\ }{\ }
      \pa_{+} c_{-}  \g_{--}  \pa_{+}^{4} \g_{--}, {\ }{\ }{\ }
      \pa_{+} c_{-}  \pa_{+} \g_{--}  \pa_{+}^{3} \g_{--},  \\[2mm]
 &
     \pa_{+}^{2} c_{-}  \g_{--} \pa_{+}^{3} \g_{--}, {\ }{\ }{\ }
     \pa_{+}^{2} c_{-}  \pa_{+} \g_{--} \pa_{+}^{2} \g_{--}, {\ }{\ }{\ }
     \pa_{+}^{3} c_{-}  \g_{--} \pa_{+}^{2} \g_{--}    \\[2mm]
 &
     \pa_{+} \g_{--}  \pa_{+}^{2} \g_{--} \pa_{+}^{3} \g_{--}, {\ }{\ }{\ }
     c_{-}\pa_{+} c_{-} \g_{--} W_{+++},    \\[2mm]
 &
    c_{-}\pa_{+} c_{-} \g_{--} \pa_{+}^{2} \f^{i}  \pa_{+} \f^{i},
    {\ }{\ }{\ }
    c_{-}\pa_{+}^{2} c_{-} \g_{--} T_{++}, \\[2mm]
 &
    c_{-}\pa_{+} c_{-}  \pa_{+} \g_{--} T_{++}, {\ }{\ }{\ }
    c_{-} \g_{--} \pa_{+} \g_{--}T_{++} T_{++}, \\[2mm]

 &  c_{-} \g_{--} \pa_{+} \g_{--}
    \pa_{+}^{2} \f^{i} \pa_{+} \f^{j} \pa_{+} \f^{k} d_{ijk}, {\ }{\ }{\ }
    \pa_{+} c_{-} \g_{--} \pa_{+} \g_{--}  W_{+++}, \\[2mm]

 &  c_{-} \g_{--} \pa_{+}^{2} \g_{--}  W_{+++}, {\ }{\ }{\ }
    c_{-} \g_{--} \pa_{+} \g_{--} \pa_{+}^{3} \f^{i} \pa_{+} \f^{i}, \\[2mm]
 &
    c_{-} \g_{--} \pa_{+} \g_{--}\pa_{+}^{2} \f^{i} \pa_{+}^{2} \f^{i},
    {\ }{\ }{\ }
    \pa_{+}^{2} c_{-}  \g_{--} \pa_{+} \g_{--} T_{++}, \\[2mm]
 &
    \pa_{+} c_{-}  \g_{--} \pa_{+}^{2} \g_{--} T_{++}, {\ }{\ }{\ }
    c_{-}  \pa_{+} \g_{--} \pa_{+}^{2} \g_{--} T_{++}, \\[2mm]

 &  \pa_{+} c_{-} \g_{--} \pa_{+} \g_{--}
    \pa_{+}^{2} \f^{i} \pa_{+} \f^{i}, {\ }{\ }{\ }
    c_{-} \g_{--} \pa_{+}^{3} \g_{--} T_{++}, \\[2mm]
 &
    c_{-} \g_{--} \pa_{+}^{2} \g_{--}
   \pa_{+}^{2} \f^{i} \pa_{+} \f^{i}, {\ }{\ }{\ }
   \g_{--} \pa_{+} \g_{--} \pa_{+}^{3} \g_{--}  T_{++},  \\[2mm]
 &
   \g_{--} \pa_{+} \g_{--} \pa_{+}^{2} \g_{--}
  \pa_{+}^{2} \f^{i} \pa_{+} \f^{i}, {\ }{\ }{\ }
  \g_{--} \pa_{+} \g_{--} \pa_{+}^{2} \g_{--} W_{+++}  \Rp \ ,
\ea\eqn{l3-subspace}
and
\eq\ba{cl}
\L^{(4)}  =  \Lp
  &   c_{-} \pa_{+} c_{-}\pa_{+}^{2} c_{-} \g_{--} b_{++}, {\ }{\ }{\ }
      c_{-}\pa_{+} c_{-} \g_{--}  \pa_{+} \g_{--}\pa_{+}^{2} b_{++}, \\[2mm]
  &   c_{-}\pa_{+}^{2} c_{-} \g_{--}  \pa_{+} \g_{--}\pa_{+} b_{++},
              {\ }{\ }{\ }
      c_{-}\pa_{+} c_{-} \g_{--}  \pa_{+}^{2} \g_{--}\pa_{+} b_{++}, \\[2mm]
  &   \pa_{+} c_{-}\pa_{+}^{2} c_{-} \g_{--}  \pa_{+} \g_{--}b_{++},
              {\ }{\ }{\ }
      c_{-}\pa_{+}^{3} c_{-} \g_{--}  \pa_{+} \g_{--} b_{++}, \\[2mm]
  &   c_{-}\pa_{+}^{2} c_{-} \g_{--}  \pa_{+}^{2} \g_{--} b_{++},
              {\ }{\ }{\ }
      c_{-}\pa_{+} c_{-}\pa_{+} \g_{--}  \pa_{+}^{2} \g_{--}b_{++}, \\[2mm]
  &
      c_{-}\pa_{+} c_{-} \g_{--}  \pa_{+}^{3} \g_{--} b_{++}, \\[2mm]
  &   c_{-}\pa_{+} c_{-} \g_{--}  \pa_{+} \g_{--}\pa_{+} \b_{+++},
              {\ }{\ }{\ }
      c_{-}\pa_{+}^{2} c_{-} \g_{--}  \pa_{+} \g_{--} \b_{+++}, \\[2mm]
  &   c_{-}\pa_{+} c_{-} \g_{--}  \pa_{+}^{2} \g_{--}\b_{+++},
              {\ }{\ }{\ }
      c_{-} \g_{--} \pa_{+} \g_{--} \pa_{+}^{2} \g_{--}\pa_{+}^{2} b_{++},
            \\[2mm]
  &   \pa_{+} c_{-} \g_{--} \pa_{+} \g_{--} \pa_{+}^{2} \g_{--}\pa_{+} b_{++},
              {\ }{\ }{\ }
      c_{-} \g_{--} \pa_{+} \g_{--} \pa_{+}^{3} \g_{--}\pa_{+} b_{++},
            \\[2mm]
 &    \pa_{+}^{2} c_{-} \g_{--} \pa_{+} \g_{--} \pa_{+}^{2} \g_{--}b_{++},
              {\ }{\ }{\ }
      \pa_{+} c_{-} \g_{--} \pa_{+} \g_{--} \pa_{+}^{3} \g_{--}b_{++},
             \\[2mm]
 &    c_{-} \g_{--} \pa_{+}^{2} \g_{--} \pa_{+}^{3} \g_{--}b_{++},
              {\ }{\ }{\ }
     c_{-} \g_{--} \pa_{+} \g_{--} \pa_{+}^{4} \g_{--}b_{++}, \\[2mm]
 &   c_{-} \g_{--} \pa_{+} \g_{--} \pa_{+}^{2} \g_{--}\pa_{+} \b_{+++},
              {\ }{\ }{\ }
     \pa_{+} c_{-} \g_{--} \pa_{+} \g_{--} \pa_{+}^{2} \g_{--}\b_{+++},
              \\[2mm]
 &   c_{-} \g_{--} \pa_{+} \g_{--} \pa_{+}^{3} \g_{--} \b_{+++},
              {\ }{\ }{\ }
     \g_{--} \pa_{+} \g_{--} \pa_{+}^{2} \g_{--} \pa_{+}^{3} \g_{--}b_{++},
               \\[2mm]
 &   c_{-}\pa_{+} c_{-} \g_{--} \pa_{+} \g_{--}b_{++} T_{++}, \\[2mm]
 &   c_{-} \g_{--} \pa_{+} \g_{--} \pa_{+}^{2} \g_{--} b_{++} T_{++}  \Rp \ .
\ea\eqn{l4-subspace}
\section{The second order consistency condition}

As we have seen in Sect. 5 the solution of the integrated
consistency condition \equ{wz} contains two nontrivial elements, given
respectively in eqs.\equ{aauu-anomaly}, \equ{aatt-anomaly}.
Let us emphasize that, actually, their numerical coefficients
turn out to be nonvanishing already at the one loop order~\cite{hl}.
Thus, according to
the Quantum Action Principle~\cite{qap}, expressions \equ{aauu-anomaly},
\equ{aatt-anomaly} yield the most general nontrivial breaking of the
Slavnov-Taylor
identity \equ{slavnov} at the order $\hbar$.
This could seem to be in contradiction with the Feynman graphs
computation of ref.~\cite{hl}, where a third term was found.
However this will be not the case, as we shall see by
discussing the consistency condition \equ{ex-rel} at the second order
(i.e. at the order $\hbar^2$). The reason relies on the fact that the third
term found in~\cite{hl} is of order $\hbar^2$, i.e. its
numerical coefficient turns out to be related to two loops Feynman diagrams.
Moreover, the presence of nonvanishing one loop anomalies implies that
the consistency condition \equ{ex-rel} at the order $\hbar^2$ is not
simply given by
the equation \equ{wz}. This means that the second order breaking terms
cannot be characterized as cohomology classes of the linearized
Slavnov-Taylor operator $\BB_{\S}$, i.e. in other words they obey a more
involved consistency condition.

For a better understanding of this point let us discuss in detail the
broken Slavnov-Taylor identity \equ{qslavnov} and the condition \equ{ex-rel} at
the order $\hbar^2$.
Let us begin with the Slavnov-Taylor identity
\eq
\SS (\G)=  \hbar \AA \cdot\G  \ .
\eqn{again-qslavnov}
Expanding $\G$ and $(\AA \cdot\G)$~\footnote{We recall that in the present
case the lowest
order of $\AA \cdot\G$ is the order $\hbar$} in powers of $\hbar$
\eq\ba{l}
  \G = \S + \hbar \G^{(1)} + \hbar^2 \G^{(2)} + O(\hbar^3) \ , \es
 \hbar \AA \cdot\G =  \hbar \AA + \hbar^2 {(  \AA \cdot\G )}^{(2)}
                    + O(\hbar^3) \ ,
\ea\eqn{hbar-expansion}
and using the identities
\eq
  \SS(\G) = \dfrac{1}{2} \BB_{\G} \G \ , \qquad
  \BB_{\FF_{1}} \FF_{2} = \BB_{\FF_{2}} \FF_{1} \ ,
\eqn{identities}
with $\FF_{1}$, $\FF_{2}$ arbitrary functionals with even ghost number,
one gets
\eq
 \BB_{\S}  \G^{(1)} = \AA  \ , \qquad  order{\ }\hbar \ ,
\eqn{slav-hbar-order}
and
\eq
 \BB_{\S}\G^{(2)} + \dfrac{1}{2} \BB_{\G^{(1)}}\G^{(1)} =
        {(  \AA \cdot\G )}^{(2)}   \ ,    \qquad  order{\ }\hbar^2 \ .
\eqn{slav-hbar2-order}
In particular, equations \equ{slav-hbar-order}, \equ{slav-hbar2-order}
show that,
while the lowest order term $\AA$ of the
breaking $(\AA \cdot\G)$ is obtained by simply computing the
$\BB_{\S}-$variation
of the one loop effective action $\G^{(1)}$, the second order term
${(  \AA \cdot\G )}^{(2)}$ requires contributions from both
$\G^{(1)}$ and $\G^{(2)}$.

Expanding now the consistency condition
\eq
   \BB_{\G} (\AA \cdot \G) = 0 \ ,
\eqn{r-ex-rel}
 one gets
\eq
  \BB_{\S} \AA = 0 \ , \qquad  order {\ }\hbar \ ,
\eqn{cons-hbar-order}
and
\eq
  \BB_{\S} {(  \AA \cdot\G )}^{(2)} + \BB_{\G^{(1)}} \AA  = 0  \ ,
   \qquad  order{\ } {\hbar^2} \ .
\eqn{cons-hbar2-order}
As it is well known, equation \equ{cons-hbar-order} implies that the lowest
nontrivial order
of $(\AA \cdot\G)$ belongs to the cohomology of $\BB_{\S}$ in the class
of the integrated local functionals with ghost number one.
Writing then the
most general solution of \equ{cons-hbar-order} as
\eq
 \AA = \AA^{(anom)} + \BB_{\S} \D  \ ,
\eqn{nontriv-sol}
$\D$ and $\AA^{(anom)}=(\AA_{\UU},\AA_{\TT})$ (see eqs. \equ{aauu-anomaly},
\equ{aatt-anomaly}) denoting respectively the trivial
and the nontrivial part of $\AA$,
equation \equ{cons-hbar2-order} becomes now
\eq
  \BB_{\S} {(  \AA \cdot\G )}^{(2)} + \BB_{\AA^{(anom)}}{\G^{(1)}}
  + \BB_{ ({\BB_{\S} \D}) } {\G^{(1)}}   = 0  \ ,
\eqn{r-cons-hbar2-order}
where use has been made of \equ{identities}.
Eq.\equ{cons-hbar2-order} is the consistency condition for
the second order breaking term ${(  \AA \cdot\G )}^{(2)}$.
One sees thus that,
in the presence of nonvanishing one loop anomalies,
${(  \AA \cdot\G )}^{(2)}$ cannot be simply detected as a cohomology class
of the operator $\BB_{\S}$. Let us conclude by recalling that the
Quantum Action Principle~\cite{qap} ensures that the right-hand side of
\equ{again-qslavnov} is the insertion of an integrated local functional
of the fields and their derivatives. This means that only the
lowest order of $(\AA \cdot\G )$ is local. Therefore the analysis of
the second order consistency condition \equ{r-cons-hbar2-order}
will, in general, require the knowledge of
the local as well as
of the nonlocal part of ${(  \AA \cdot\G )}^{(2)}$.


\end{document}